\documentclass[aps,pra,reprint]{revtex4-1}

\usepackage{amsmath}
\usepackage{graphicx}
\usepackage{dcolumn}
\usepackage{verbatim}

\newcommand{\spt}[1]{$\sp{\textrm{#1}}$}
\newcommand{\rhon}[1]{\rho^{(#1)}}
\newcommand{\nuks}{\nu_{ks}}
\newcommand{\nuksr}{\nu_{ks}(\mathbf{r})}
\newcommand{\nuksnn}[1]{\nu_{ks}^{(#1)}}
\newcommand{\nuksn}[1]{\nu_{ks}^{(#1)}(\mathbf{r})}
\newcommand{\ppnks}{\frac{\partial\nu_{ks}}{\partial\rho}}

\newcommand{\br}{\mathbf{r}}
\newcommand{\bU}{\mathbf{U}}

\begin{document}

\title{General degeneracy in density functional perturbation theory}
\author{Mark C. Palenik}
\thanks{NRC Research Associate}
\email{mark.palenik.ctr@nrl.navy.mil}
\author{Brett I. Dunlap}
\affiliation{Code 6189, Chemistry Division, Naval Research
Laboratory, Washington, DC 20375, United States}

\begin{abstract}
Degenerate perturbation theory from quantum mechanics is inadequate in density functional theory (DFT) because of nonlinearity in the Kohn-Sham potential. Herein, we develop the fully general perturbation theory for open-shell, degenerate systems in Kohn-Sham DFT, without assuming the presence of symmetry or equal occupation of degenerate orbitals. To demonstrate the resulting methodology, we apply it to the iron atom in the central field approximation, perturbed by an electric quadrupole. This system was chosen because it displays both symmetry required degeneracy, between the five 3\textit{d} orbitals, as well as accidental degeneracy, between the 3\textit{d} and 4\textit{s} orbitals. The quadrupole potential couples the degenerate 3\textit{d} and 4\textit{s} states, serving as an example of the most general perturbation.
\end{abstract}

\maketitle

\section{Introduction}

When a small, perturbing potential is applied to a system with degeneracy, quantum mechanics tells us that the the eigenvalues typically split.  The first-order splitting can be found by diagonalizing the perturbing potential, $V^{(1)}$, within the degenerate subspace \cite{sakurai2011modern}.

In contrast, we \cite{Palenik2016} and others \cite{Cances2014} have proven that for symmetry required degeneracy in open shell systems, the eigenvalues in Kohn-Sham (KS) \cite{Kohn1965} density functional theory (DFT) \cite{HKTheorems} do not split.  KS DFT allows for fractional occupation numbers at the Fermi-level, and the application of a small, symmetry-breaking potential induces a fractional transfer of Fermi-level electrons that restores the original degeneracy.

Herein, we will develop the perturbation theory that also applies to systems with accidental degeneracy.  We impose no assumptions about the occupation numbers of the initial state or its symmetry.  We only note that some, but not necessarily all, degenerate orbitals may be equally occupied.

The resulting methodology is not simply the application of coupled-perturbed Kohn-Sham (CPKS) \cite{Komornicki1993,Gonze1997,Fournier1990} to the equations from degenerate, quantum mechanical perturbation theory.  Solving for the perturbed orbitals in CPKS requires the inversion of a large matrix with dimension equal to the product of the number of occupied and virtual orbitals, which is typically done through an iterative procedure.  This is due to the nonlinearity of the Coulomb and exchange-correlation (XC) potentials, which we collectively refer to as $\nuks$. However, careful analysis of the degenerate problem shows that the KS-DFT solution requires perturbed occupation numbers in addition to perturbed orbitals, which is not a feature of CPKS.

Perturbation theory allows for electronic states and energies to be differentiated with respect to some change in the Hamiltonian.  In molecular physics, it can be used to differentiate the energy with respect to nuclear coordinates and to find vibrational modes.  Along these lines, Jahn-Teller distortions can be understood by applying perturbation theory to a degenerate state, which results in a Taylor series of the energy with respect to nuclear displacements \cite{Jahn1937}.  Despite the importance of the Jahn-Teller effect we are aware of no commercial computer codes capable of treating it as a perturbation.

As we know of no Jahn-Teller system with accidental degeneracy, we demonstrate the applicable, general perturbation theory using the fractionally-occupied atomic iron ground state in the central field approximation.  Its equally occupied 3\textit{d} orbitals are accidentally degenerate with the differently occupied 4\textit{s} orbital.  A quadrupole electric field can couple all of the degenerate states, demonstrating the differences between the treatment of the symmetry required and accidental degeneracies.  Although the initial SCF solution for the iron atom is performed in spherical symmetry, this is not a requirement of the general theory we outline in section~\ref{SecTheory}.

Symmetry required degeneracy in KS DFT only occurs when the orbitals of a given spin within each irreducible representation of the symmetry group are equally occupied, giving $\nuks$ the same symmetry as the external potential.  For the specific case of equally occupied Fermi-level orbitals, we previously generated a perturbation series using an imaginary-time propagator in the limit that $t\rightarrow\infty$, normalized to preserve the number of Fermi-level electrons \cite{Palenik2016}.  This is equivalent to the zero-temperature limit of a thermal state and selects the lowest lying eigenstate meeting the normalization conditions.

We will now do the same for systems where the degenerate orbitals do not have to be equally occupied, and prove that the lack of eigenvalue splitting is not unique to symmetry required degeneracy.  We will then show that the resulting equations can always be solved, provided that the Coulomb and XC potentials are a function of the density, by a change in occupation numbers combined with a series of orthogonal transformations between groups of degenerate orbitals.  Next, we will apply the resulting methodology to atomic iron.  Conclusions follow.

\section{Theory}
\label{SecTheory}

Consider a system of electrons with a degenerate Fermi level where the Fermi-level orbitals can each have any occupation numbers between zero and one.  If the system is open-shell, there can be non-trivial orbital mixing and electron-transfer within the degenerate space.

The nonlinearity of KS DFT means that under a perturbation, these Fermi-level occupation numbers must necessarily change.  We previously showed that this is true for the specific case when the initial occupation numbers are equal \cite{Palenik2016}.  As we noted, when the degenerate states start out equally occupied, $V^{(1)}$ can be diagonalized without changing the unperturbed density.  However, this is not sufficient to solve the perturbation problem, because $V^{(1)}$ will cause the previously degenerate levels to split, resulting in fractionally occupied excited states.  Therefore, to find the ground state, the occupation numbers must be allowed to change.

Now, we note that the unperturbed, accidentally degenerate orbitals need not be equally occupied.  If, for example, two degenerate orbitals have different occupation numbers, a unitary transformation cannot be made between them without changing the density, which invalidates the original self-consistent field (SCF) calculation.  Therefore, the perturbing potential cannot be diagonalized, and without diagonalization, the initial state will evolve into a linear combination of states with different eigenvalues.  We will show that once again, the remedy to this problem involves changing the fractional occupation numbers.

Perturbation theory builds a differentiable map from a state in the unperturbed system to an adiabatically connected state in the perturbed system.  We wish to build a map between the ground states.  Because this map is differentiable, quantities such as occupation numbers and orbital rotation matrices must change continuously with the strength of the perturbing potential.  The only way to build such a map in KS DFT is by conserving the number of electrons at the Fermi-level.

A general perturbation will cause a degenerate Fermi-level to split.  To ensure that the system remains in the ground state, we must transfer electron density into the lowest available state that is not fully occupied.  In the standard formulation of quantum mechanics, this is equivalent to diagonalizing the perturbing potential and filling the lowest states in order.  However, in KS DFT, as we begin to rotate the degenerate orbitals into one another and transfer electrons into the lowest possible states, the eigenvalues simultaneously shift.

Orbital eigenvalues are the derivatives of the energy with respect to occupation numbers \cite{Janak1978}.  Although in the exact functional as defined by Perdew \cite{Perdew1982,Sagvolden2008}, eigenvalues are constant, except for discontinuous shifts at integer occupation numbers, this behavior is not shared by existing, commonly used approximate functionals.  Typical expressions for the XC energy-density are continuous, differentiable functions of the electron-density.  Furthermore, in the exact functional as defined by Cohen and Wasserman, eigenvalues change continuously as a function of occupation number \cite{Cohen2003}, and this is also a property of Fermi liquids \cite{landau1957,Landau1959,Thouless1972quantum,Nesbet1997}.

The derivative of the eigenvalues with respect to occupation numbers is the Hessian of the energy.  This matrix determines how the eigenvalues change as electrons are transferred between orbitals.  If the unperturbed state is the ground state, it is an energy minimum, meaning that the Hessian of the energy is positive definite.  Therefore, an eigenvalue derived from a fully variational SCF calculation will increase as the occupation number of the corresponding orbital is increased.

In addition to its own occupation number, the eigenvalue of a given orbital depends on the occupation numbers of all other orbitals.  If the number of electrons is to be conserved, electrons that are added to one orbital must be removed from another.  Therefore, the off-diagonal elements of the Hessian also play a role in the behavior of the eigenvalues.

Still, we can paint a qualitative picture of degenerate perturbation theory as a two step process.  First, a perturbing potential is applied, which causes the degenerate levels to split.  Next, electrons are transferred from orbitals with higher eigenvalues to those with lower eigenvalues until degeneracy is restored, at which point no more electrons can be transferred.  Because we are not requiring the degenerate orbitals to be equally occupied, we will show that orbital rotations within the degenerate space will also be required at first-order and higher to zero the off-diagonal matrix elements.  This also has an impact on the eigenvalues.

We can additionally think of the fact that the number of Fermi-level electrons is held constant in another way.  The denominator that results in this normalization is the only possible choice that causes the fractionally occupied excited states produced by Fermi-level splitting to vanish, while scaling the remainder so that the total number of electrons is conserved.  Conserving the number of electrons is necessary in DFT perturbation theory because of the nonlinearity of $\nuks$.  Typically, in quantum mechanical perturbation theory, the intermediate normalization is often used, which makes the overlap of the perturbed and unperturbed orbitals equal to unity, but this causes the integral of the perturbed electron density to change, which would affect the computation of the Coulomb and XC energies if it were used.

We can build a properly normalized, differentiable map between the unperturbed and perturbed states by writing each Fermi-level orbital $\phi_i$ as a function of $\lambda$, which scales the strength of the perturbing potential.  We again use an imaginary-time propagator in the limit that $t\rightarrow\infty$ \cite{Palenik2016}. This is most easily done in the interaction picture, where the perturbed KS potential is given by
\begin{equation}
    H'_{IP} = e^{-H_{ks}t}\left(\lambda V^{(1)} + \sum_{n=1}^{\infty}\lambda^{n}\nuksnn{n}\right)e^{H_{ks}t}.
\end{equation}
Here, $H_{ks}$ is the unperturbed operator from the KS equation $H_{ks}\phi_i=\epsilon_i\phi_i$.  The Coulomb and XC potentials, because they are functions of the density, change at all orders of perturbation theory, leading to the infinite sum.  Our interaction picture imaginary-time propagator can then be written as
\begin{equation}
|\phi_i(\lambda)\rangle=\frac{\sum_k|\phi_k\rangle\langle\phi_k|\mathcal{T}e^{-\int_0^\infty
H'_{IP}dt}|\phi_i\rangle}{\sum_m \frac{n_m(\lambda)}{N_e} \sqrt{\sum_j|\langle\phi_j|\mathcal{T}e^{-\int_0^\infty
H'_{IP}dt}|\phi_m\rangle|^2}},
\label{PhiGround}
\end{equation}
where $\mathcal{T}$ is the time ordering operator, $n_m(\lambda)$ is an occupation number as a function of $\lambda$, and $N_e$ is the number of Fermi-level electrons.  The index $m$ runs over the degenerate orbitals, and $j$ and $k$ run over all orbitals at or above the Fermi level.  

This normalization is nearly identical to the one we used in our derivation for equally occupied orbitals, except that the factor of $1/N_d$, where $N_d$ is the number of degenerate orbitals, has been replaced by $n_m(\lambda)/N_e$.  The normalization we have chosen here ensures that $\sum_m n_m(\lambda)\langle\phi_m^*(\lambda)|\phi_m(\lambda)\rangle = N_e$, thus conserving the number of electrons.  The old normalization only conserves the number of electrons if each Fermi-level orbital is equally occupied.

In addition to holding the number of electrons constant, we can see that the denominator will cause any excited states in the numerator to vanish.  In the limit that $t\rightarrow\infty$, the denominator goes as $e^{-\epsilon(\lambda)t}$, where $\epsilon(\lambda)$ is the smallest eigenvalue adiabatically connected to any of the Fermi orbitals.  The numerator goes as $e^{-\epsilon_i(\lambda)t}$, which is the eigenvalue of the state adiabatically connected to $\phi_i(\lambda)$. Therefore, the entire equation goes as $e^{[\epsilon(\lambda)-\epsilon_i(\lambda)]t}$, which approaches zero as $t\rightarrow\infty$ if $\epsilon_i(\lambda)>\epsilon(\lambda)$.

The first-order orbitals are the first term in the Taylor series of $\phi_i(\lambda)$ at $\lambda=0$.  Therefore, we need to differentiate Eq.~(\ref{PhiGround}) once.  When there is no degeneracy, this reproduces the standard RSPT sum over states expression.  We define degenerate perturbation theory by imposing the requirement that the matrix elements between degenerate orbitals approach a well defined value as $t\rightarrow\infty$. We do this by setting the time-derivative of these matrix elements to zero in that limit.  For degenerate orbitals $\phi_k$ and $\phi_i$ at first order, this yields
\begin{equation}
\langle\phi_k|V^{(1)}+\nuksnn{1}|\phi_i\rangle=
\delta_{ik}\sum_m \frac{n_m}{N_e}\langle\phi_m|V^{(1)}+\nuksnn{1}|\phi_m\rangle.
\label{EqFirstOrder}
\end{equation}

As we proved for equal occupation numbers, Eq.~(\ref{EqFirstOrder}) says that the first-order potential is diagonal within the degenerate subspace and that the first-order eigenvalues are identical.  Because all of the diagonal elements are identical, the weighting factors of $n_m/N_e$ on the right hand side can be replaced with $1/N_d$, arriving at
\begin{equation}
\langle\phi_k|V^{(1)}+\nuksnn{1}|\phi_i\rangle=
\frac{\delta_{ik}}{N_d}\sum_m \langle\phi_m|V^{(1)}+\nuksnn{1}|\phi_m\rangle,
\label{EqFirstOrder2}
\end{equation}
which is exactly the same result we found for symmetry required degeneracy \cite{Palenik2016}.

The non-linear $\nuks$ operator is responsible for the fundamental differences between the behavior the KS DFT model system and the quantum mechanical wave function.  As long as it is present, the eigenvalues can be made equal by adjusting occupation numbers and orbital rotations within the degenerate space.  The change in occupation numbers required to restore degeneracy when a perturbation is applied reduces the symmetry of the unperturbed electron density.  Because the eigenvalues depend on the occupation numbers, this also means that despite the presence of degeneracy, the perturbed state is unique \cite{Cances2014,Palenik2016}.  Attempting to adjust the occupation numbers away from the values prescribed by the perturbation series breaks the eigenvalue degeneracy and moves the system away from the ground state.

If we take the limit of a noninteracting system by scaling $\nuks$ with a small constant, the first-order occupation numbers and orbital rotations will grow outside of the range of physically reasonable values, approaching infinity as $\nuks$ goes to zero.  To see this, we note that the first-order density in Eq.~(\ref{EqFirstOrder2}) is implicitly contained in $\nuksnn{1}$, which is defined as
\begin{equation}
\nuksn{1}=\int \frac{\delta\nuksr}{\delta\rho(\br')}\rhon{1}(\br')d\br'\equiv\ppnks\rhon{1}.
\label{nuks1}
\end{equation}
The variation of $\nuks$ with respect to $\rho$ is the Hessian of the electron-electron interaction energy.  Although we will more rigorously solve Eq.~(\ref{EqFirstOrder2}) in Section~\ref{SectionIron}, we can note that if we rearrange its terms to solve for the matrix elements of $\rhon{1}$, they will be proportional to the inverse of this Hessian.

If the electron-electron interaction is scaled by a constant, $k$, the inverse of its Hessian will go as $1/k$.  At $k=0$, the Hessian will become singular and the matrix elements of $\rhon{1}$ within the degenerate space will become infinite \cite{Palenik2016}.

The first-order density is given by
\begin{equation}
\rhon{1} = \sum_{i}\left(n_i^{(1)}\phi^*_i\phi_i+n_i\sum_j2\mathrm{Re}\bU^{(1)}_{ij}\phi^*_i\phi_j\right),
\label{EqRho1}
\end{equation}
where $n_i$ is the occupation number of orbital $i$ and a lack of a superscript indicates an unperturbed quantity.  The indices $i$ and $j$ run over all orbitals, but $n^{(1)}_i$ is only nonzero at the Fermi level.  The matrix $\bU^{(1)}$ is the first-order unitary transformation that mixes orbitals.  Therefore, if the matrix elements of $\rhon{1}$ become infinite, the quantities $\bU^{(1)}_{ij}$ and $n_i^{(1)}$ that define them must become infinite as well.

Qualitatively, this means that a weaker electron-electron interaction requires a larger shift in electron-density to restore the degeneracy that is broken by the external perturbing potential.  For a given perturbation, the required change in occupation numbers may fall outside the limits imposed by Fermi statistics.  At this point, the occupation numbers will cease to change, whether degeneracy has been restored or not.  However, as $\lambda$ is made smaller, the change in occupation numbers becomes smaller as well.  Therefore, we should expect this perturbation series to provide a physically meaningful mapping between states as long as the perturbation is weak enough that Fermi statistics are obeyed.

The perturbed orbitals, occupation numbers, and eigenvalues were all generated by taking their derivatives with respect to the strength of the perturbing potential.  These quantities can, then, always be interpreted as derivatives at $\lambda=0$, regardless of the strength of the perturbation.  Fermi statistics only limits the range of $\lambda$ over which these derivatives can be used to form a Taylor series connecting the perturbed and unperturbed states.  At the point where one or more occupation numbers can no longer change, a derivative discontinuity will occur.

The tendency of the first-order orbital rotations and occupation numbers toward infinity as the equations approach linearity is a function of the fact that the differentiability of the unperturbed state is lost.  For the true wave function, which is an eigenstate of a linear operator, an arbitrary initial state cannot be perturbed into an eigenstate.  Rather, only appropriate linear combinations of the unperturbed, degenerate eigenstates correspond to eigenstates of the perturbed system.  It is the nonlinearity of $\nuks$ that guarantees the differentiability of the initial state.

We will now show that Eq.~(\ref{EqFirstOrder2}) can be solved as long as $\nuks$ depends on the electron-density.  The first-order perturbing potential is a Hermitian operator.  A real, Hermitian operator has $N_d(N_d-1)/2$ off-diagonal elements in the degenerate subspace.  Zeroing them results in an equal number of constraint equations.  The requirement that the diagonal elements are equal produces another $N_d-1$ constraints \cite{Ullrich2002}.

Each constraint requires one adjustable parameter for the equations to be solvable.  An orthogonal transformation can be written as the exponential of an antisymmetric matrix, which provides $N_d(N_d-1)/2$ adjustable parameters.  The first-order Fermi-level occupation numbers, which must sum to zero \cite{Palenik2016}, provide $N_d-1$.  Therefore, an orthogonal transformation between degenerate orbitals combined with a change in first-order occupation numbers provides the appropriate number of parameters for solving first-order perturbation theory.  For a Hermitian $V^{(1)}$ with imaginary components, a unitary, rather than orthogonal, transformation results in the required number of parameters.

Unfortunately, it is not possible to make an arbitrary unitary transformation within the degenerate space in KS DFT, like it is in standard quantum mechanics.  Unless the unperturbed orbitals are equally occupied, applying a unitary transformation to the unperturbed system would cause the electron-density to change, thereby requiring a new SCF solution.  Therefore, we can only use orbital rotations to diagonalize the first-order potential within each group of equally occupied degenerate orbitals.  If the off-diagonal elements that couple orbitals with \textit{different} occupation numbers are to be zeroed, it must be because the matrix elements of $\nuksnn{1}$ cancel those of $V^{(1)}$.

Because $\nuksnn{1}$ depends on $\rhon{1}$, from Eq.~(\ref{EqRho1}), the two sets of adjustable parameters that can potentially affect its matrix elements are $n^{(1)}_i$ and $\bU^{(1)}_{ij}$.  Therefore, if it is possible to solve Eq.~(\ref{EqFirstOrder2}), we must be able to show that the matrix $\bU^{(1)}$ has exactly the right number of adjustable parameters to cancel the remaining off-diagonal elements of $V^{(1)}$.

We can begin by dividing the matrix $\bU^{(1)}$ into two parts: one that mixes orbitals with different eigenvalues and one that acts within the degenerate subspace.  Between nondegenerate orbitals, the matrix elements of $\bU^{(1)}$ are given by the usual Rayleigh-Schr\"odinger perturbation theory sum over states \cite{Shavitt2009}. Between equally occupied degenerate orbitals, $\bU^{(1)}$ can be neglected until second order, because the elements of $\bU^{(N)}$ that mix equally occupied orbitals do not affect the density until order $N+1$ \cite{Palenik2016}.

However, between orbitals with different occupation numbers, $\bU^{(1)}$ does have an effect on the density, and therefore, on the matrix elements of $\nuksnn{1}$.  Combining the total number of parameters involved in a series of orthogonal transformations within each set $S$ of equally occupied orbitals (typically a particular irreducible representation of the symmetry group) and the elements of $\bU^{(1)}$ that mix orbitals from the sets $S$ and $S'$, which have different occupation numbers (typically, different representations of the symmetry group), the total number of adjustable parameters in $\bU^{(1)}$ is
\begin{equation}
\sum_S\mkern-5mu \left[\frac{1}{2}N_S\left(N_S-1\right)+\mkern-5mu\sum_{S'>S}N_S N_{S'}\right]\mkern-5mu =\mkern-2mu \frac{1}{2}N_d\left(N_d-1\right).
\end{equation}
Adding the $N_d-1$ first-order occupation numbers, we now have enough variables to solve the first-order perturbation equations.  By the same arguments, it can be shown that at $N$th order, $\bU^{(N-1)}$ between equally occupied orbitals, $\bU^{(N)}$ between orbitals with different occupation numbers, and $n_i^{(N)}$ provide the required number of parameters to solve the equations.

Substituting the definitions of $\nuksnn{1}$ and $\rhon{1}$ from Eq.~(\ref{nuks1}) and Eq.~(\ref{EqRho1}) into Eq.~(\ref{EqFirstOrder2}) leads to a system of equations for the first-order occupation numbers, first-order mixing between degenerate states, and zeroth-order transformations between equally occupied states.  
With these substitutions, we can rearrange Eq.~(\ref{EqFirstOrder2}) to get:
\begin{equation}
\begin{split}
\langle\phi_k|\ppnks\sum_{i}\left(n_i^{(1)}\phi^*_i\phi_i+n_i\sum_j2\mathrm{Re}\bU^{(1)}_{ij}\phi^*_i\phi_j\right)|\phi_i\rangle\\
= \epsilon^{(1)}\delta_{ik}-\langle\phi_k|V^{(1)}|\phi_i\rangle,
\end{split}
\label{EqFirstOrderFull}
\end{equation}
where $\epsilon^{(1)}$ is a constant equal to $1/N_d$ times the trace of the entire first-order potential in the degenerate space
\begin{equation}
    \epsilon^{(1)} = \frac{1}{N_d}\sum_m \langle\phi_m|V^{(1)}+\nuksnn{1}|\phi_m\rangle.
    \label{EqEps1}
\end{equation}

\section{Atomic Iron}
\label{SectionIron}

To explore Eq.~(\ref{EqFirstOrderFull}) further, we will turn to the VWN \cite{Vosko1980} ground state of atomic iron, which has 3$d$-4$s$ degeneracy \cite{Averill1992}.  Degeneracy between the $d$ states is due to spherical symmetry, whereas the $s$-$d$ degeneracy is accidental.  Our SCF calculations found fractional occupation numbers of 0.299 in all five 3$d$ spin up states and 0.504 in the 4$s$ spin up state.

Complex spherical harmonics, $Y_\ell^m$, were used for the angular portion of the orbitals. Because we are mainly interested in the behavior of the degenerate space, we assume that excitations into virtual orbitals can be ignored, which allows us to avoid the computational complexities introduced by occupied-virtual mixing \cite{Komornicki1993,Gonze1997,Fournier1990,Palenik2015}.  We can then uniquely identify the orbitals by the labels $\phi_\ell^m$, corresponding to angular momentum $\ell$ ($0$ or $2$) and azimuthal component $m$ (ranging from $-\ell$ to $\ell$).  To couple degenerate states, the perturbing quadrupole potential $V^{(1)}=\lambda Y_2^0/|\br|^3$ was introduced.  This potential is diagonal and traceless within the $d$ states, has an $s$ matrix element of zero, and couples $\phi_2^0$ with $\phi_0^0$.

Because the unperturbed occupation numbers depend only on the total angular momentum, we will refer to them by a single index, $n_\ell$.  The first-order occupation numbers depend on $\ell$ and $m$, and therefore will be labeled with two indices, $n_\ell^{m(1)}$.

For this specific problem, we can make several simplifications to Eq.~(\ref{EqFirstOrderFull}) based on symmetries.  The first term on the left-hand side, in our current notation is
\begin{equation}
\langle\phi_{\ell}^k|\sum_{l,m}n^{m(1)}_l\phi_l^{m*}\phi_l^m|\phi_{\ell'}^i\rangle.
\end{equation}
Because both $\phi^{m*}_l$ and $\phi_l^m$ have the same azimuthal component of angular momentum, their matrix elements must be diagonal in the azimuthal index.

In our current notation, $\epsilon^{(1)}$, which appears on the right-hand side of Eq.~(\ref{EqFirstOrderFull}), is given by
\begin{equation}
    \epsilon^{(1)}=\sum_{lm}n_l\langle\phi_l^m|V^{(1)}+\ppnks\rhon{1}|\phi_l^m\rangle.
    \label{Epsilon12}
\end{equation}
We can simplify this expression by noting that the unperturbed density is spherically symmetric, meaning that $\nuks$ and its derivatives (e.g. $\ppnks$) are spherically symmetric as well.  Summing the product of $\phi_l^{*m}\phi_l^m$ over $m$ also results in a spherical function.  The second term on the right hand side of Eq.~(\ref{Epsilon12}) is, then, the integral of $\rhon{1}$ with a spherically symmetric function.  This allows us to utilize the orthogonality of the spherical harmonics that appear within $\rhon{1}$ itself. 

The second term in $\rhon{1}$ from Eq.~(\ref{EqRho1}) comes from the first-order unitary transformation within the degenerate space. In our current notation, it is written as
\begin{equation}
\bU^{mm'(1)}_{\ell\ell'}\phi_\ell^{m*}\phi_{\ell'}^{m'}.
\end{equation}
Orthogonality of spherical harmonics means that, when integrated with a spherical function, this term is zero unless $\ell,m=\ell',m'$.  However, these terms are also zero because of the antisymmetry of $\bU^{(1)}$.  This can be shown by noting that a unitary transformation can be written as the exponential of $i$ times a Hermitian matrix, $\mathbf{M}$
\begin{equation}
\bU = e^{i\mathbf{M}} = 1 + i\mathbf{M} +\ldots.
\end{equation}
From this expansion, it is clear that $\bU^{(1)}$ is anti-hermitian.  Any diagonal elements of $\bU^{(1)}$ are necessarily imaginary.  Because the perturbing potential is an electric quadrupole, which respects time symmetry, $\bU$ can be written as a real, orthogonal transformation.  Therefore, $\bU^{(1)}$ is antisymmetric, and so, its contribution to $\epsilon^{(1)}$ vanishes.

Finally, we can look at the matrix elements involving $\bU^{(1)}$ on the left-hand side.  Because $\bU^{(1)}$ is antisymmetric, there is cancellation between any terms in $\sum_{\ell m,\ell'm'} n_\ell\bU^{mm'(1)}_{\ell\ell'}\phi^{m*}_\ell\phi_{\ell'}^{m'}$ where $n_\ell$ is equal to $n_{\ell'}$ .  Therefore, $\bU^{(1)}$ can only mix $d$ and $s$ states, which have different occupation numbers.  Furthermore, angular momentum addition rules limit the terms that contribute to each matrix element. Applying all of these simplifications to Eq.~(\ref{EqFirstOrderFull}), we get
\begin{equation}
\begin{split}
\langle\phi_\ell^k|V^{(1)}|\phi_{\ell'}^{i}\rangle+\sum_{l,m}n_l^{m(1)}\langle\phi_\ell^k|\ppnks\phi_l^{m*}\phi_l^m|\phi_{\ell'}^{k}\rangle\delta_{ik}\\
+2(n_2-n_0)\bU^{\mu 0(1)}_{20}\langle\phi_\ell^k|\ppnks\mathrm{Re}\phi_2^{\mu *}\phi_0^0|\phi_{\ell'}^i\rangle
=\frac{\delta_{ik}\delta_{\ell\ell'}}{N_d}\\
\times\sum_{l,m}\langle\phi_l^m|V^{(1)}+\ppnks\sum_{l',m'}n_{l'}^{m'(1)}\phi^{m'*}_{l'}\phi^{m'}_{l'}|\phi_l^m\rangle,
\end{split}
\label{EqFirstOrderFinal}
\end{equation}
where $\mu$ is equal to $i-k$, because angular momentum addition rules specify that only these four-orbital integrals are nonzero.  The $\bU^{(1)}$ term should additionally be taken to be zero if $i-k$ is outside of the range -2 to 2.

A first step in solving Eq~(\ref{EqFirstOrderFinal}) is to apply an orthogonal transformation that diagonalizes the left-hand side along the $m$ index when $\ell=\ell'$.  The second term, containing a factor of $n_l^{m(1)}$, will always be diagonal when $\ell=\ell'$ under any orthogonal transformation that mixes azimuthal components of angular momentum, due to group symmetries.  Therefore, it is only the first and third terms that need to be diagonalized.

The potential $V^{(1)}$ has an angular factor of $Y_2^0$, and therefore, couples only $\phi_0^0$ and $\phi_2^0$, which is the same set of orbitals coupled by the $n_l^{m(1)}$ and the $\bU_{20}^{00(1)}$ terms.  Therefore, the only nonzero component of the first-order unitary transformation needed is $\bU_{20}^{00(1)}$.  The five independent first-order occupation numbers and the single independent component of $\bU^{(1)}$ provide the six adjustable parameters needed to equate the diagonal elements of Eq.~(\ref{EqFirstOrderFinal}) and zero the single off-diagonal element.

To solve Eq.~(\ref{EqFirstOrderFinal}), we implemented a variable metric method to minimize the sum of the squares of the left-hand side minus the right-hand side.  The resulting parameters, rounded to three significant figures, are given in Table~\ref{TableFO}.

\begin{table}[t]
	\caption{Nonzero parameters from the first-order density}
	\centering
	\begin{ruledtabular}
		\begin{tabular}{r|c c c c c c}
		Parameter&$\bU_{20}^{00(1)}$&$n^{\pm2(1)}_{2}$&$n^{\pm1(1)}_{2}$&$n^{0(1)}_{2}$&E\spt{(2)}&E\spt{(3)}\\
		Value& 0.455& 11.9& -5.94& -11.9& -16.2& 8.55$\times$10\spt{-3}\\
		\end{tabular}
	\end{ruledtabular}
	\label{TableFO}
\end{table}

\begin{figure*}[t]
	\includegraphics[width=0.5\columnwidth]{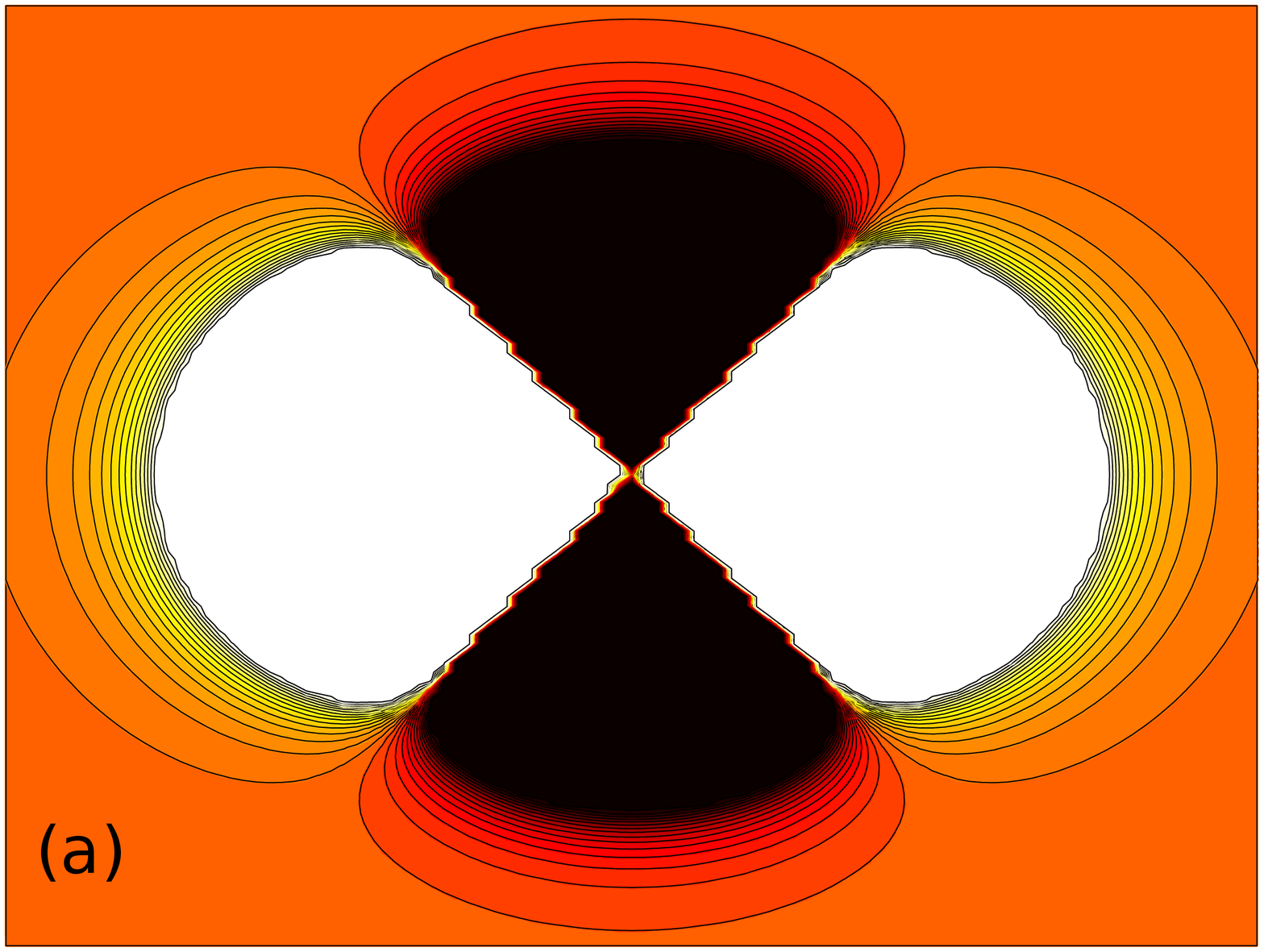}
	\includegraphics[width=0.5\columnwidth]{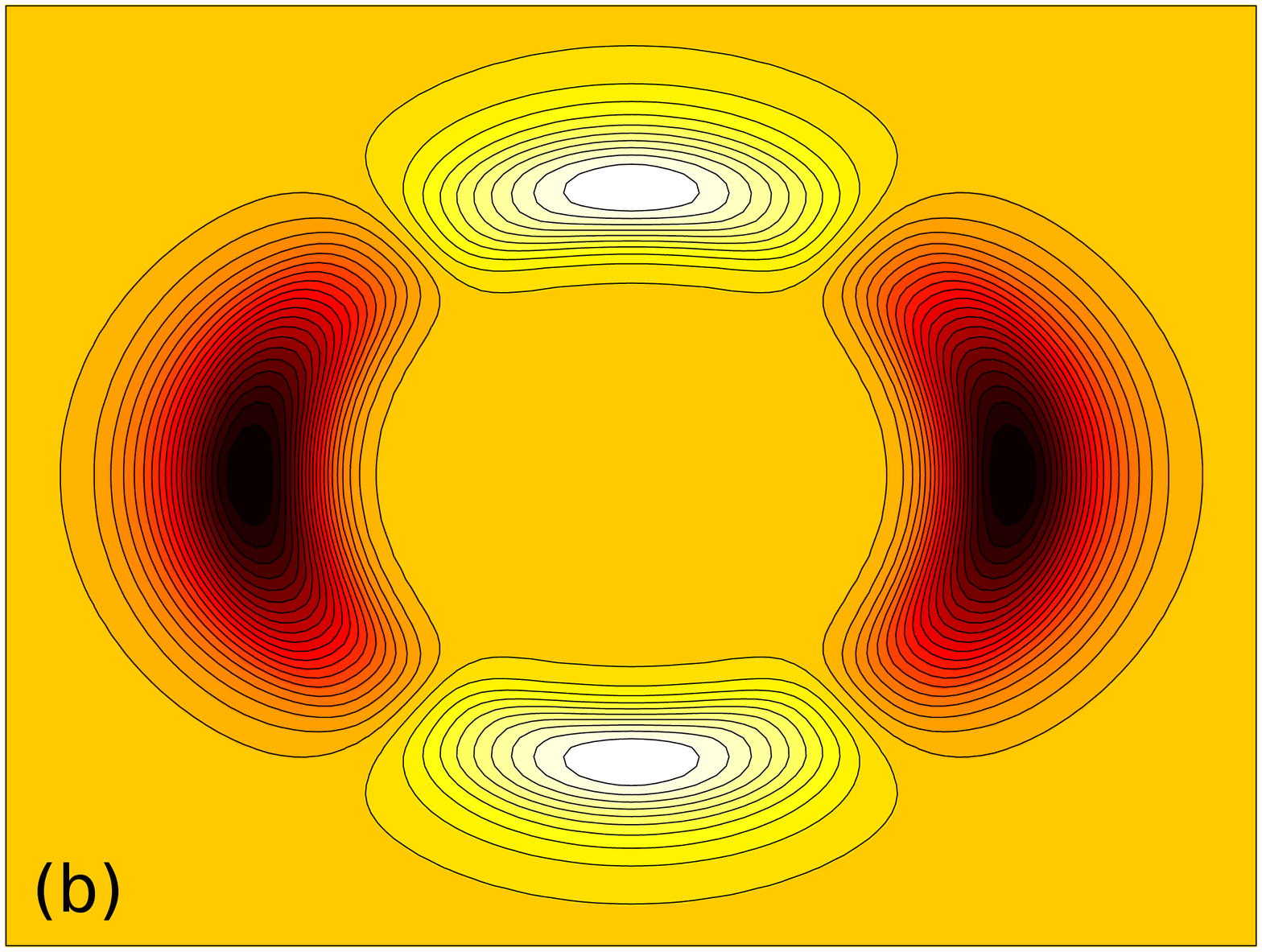}
	\includegraphics[width=0.5\columnwidth]{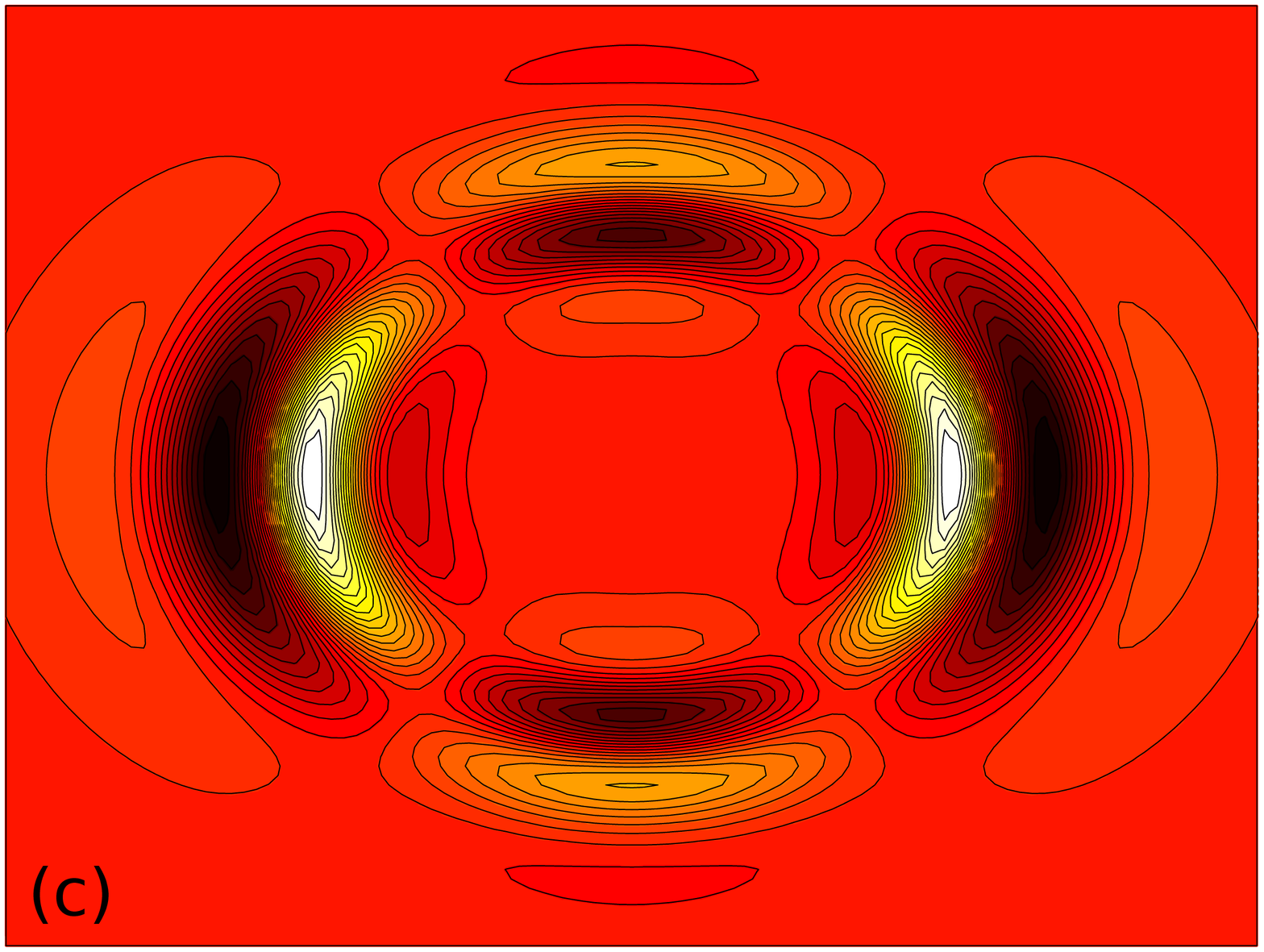}
	\includegraphics[width=0.5\columnwidth]{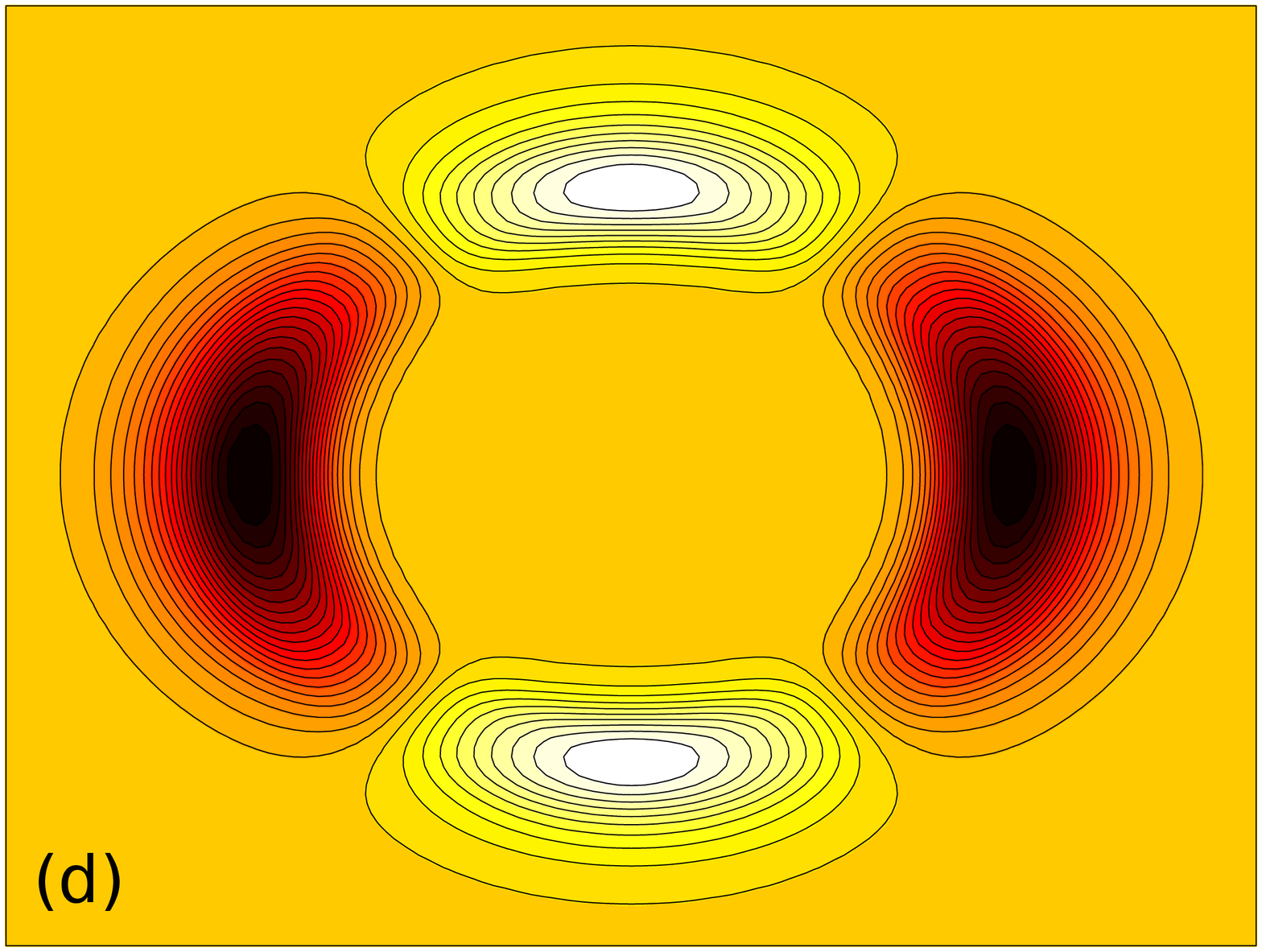}
	\caption{(a) perturbing external potential $V^{(1)}$ (b) first-order density due to change in occupation numbers (c) first-order density due to $s$-$d$ mixing (d) total first-order density.  Lighter colors represent higher values and darker colors are lower.}
	\label{figrhoandv}
\end{figure*}

The first-order density can be used to find the first through third-order energies \cite{Wigner1935,Angyan2009,Cances2014}.  In terms of $\rhon{1}$, the general expressions are the same as when the occupation numbers are equal \cite{Palenik2016}: 
\begin{eqnarray}
&E^{(1)}& = \sum_i n_i^{(0)}\langle\phi_i|V^{(1)}+\nuksnn{1}|\phi_i\rangle\\
&E^{(2)}& = -\frac{1}{2}\int\frac{\delta\nuksr}{\delta\rho(\br')}\rhon{1}(\br)\rhon{1}(\br')\\\
&E^{(3)}& = -\frac{1}{6}\int\frac{\delta^2\nuksr}{\delta\rho(\br')\delta\rho(\br'')}\rhon{1}(\br)\rhon{1}(\br')\rhon{1}(\br'').
\end{eqnarray}
$E^{(1)}$ is zero because the trace of the perturbing potential is zero.  The second-order energy was found to be -16.2~hartrees, and the third-order energy is 8.55$\times$10\spt{-3}~hartrees.  The comparatively small value of $E^{(3)}$ is due to the fact that the contributions from terms containing three factors of the first-order occupation numbers sum to zero.  All contributing terms have at least one factor of the much smaller $\bU_{20}^{00(1)}$.  $E^{(3)}$ is only nonzero because $V^{(1)}$ mixes the $s$ and $d$ states.

Figure~\ref{figrhoandv} contains plots of the perturbing potential and first-order density along the radial and $\theta$ axes ($\theta$ corresponding to the axis for which $Y_\ell^m$ is independent of $m$).  The radial axis is plotted along points used in the parameter-free 80-point quasi-experimental radial integration grid of K\"oster et al. \cite{Koester2004-2} and does not scale linearly with distance from the center of the plot.  The colors are relative to the maximum and minimum values of each individual plot.

In Fig.~\ref{figrhoandv}(b), we have isolated the portion of the first-order density produced by the first-order occupation numbers.  In Fig.~\ref{figrhoandv}(c), we have plotted the portion of the first-order density that is caused by mixing of the $\phi_2^0$ and $\phi_0^0$ orbitals.  The total density, in (d), is nearly identical to (b) because the density in (b) is, on average, three orders of magnitude larger than in (c).

The first-order occupation numbers in Table~\ref{TableFO} are one to two orders of magnitude larger than the individual $d$ orbital occupation numbers.  However, they scale linearly with $\lambda$, which can be made arbitrarily small as the external perturbing field is made weaker.  At the point where the perturbing field becomes strong enough that electrons can no longer be transferred between degenerate orbitals, the degeneracy will cease.

\section{Conclusions}

We have generalized our previous work on degenerate perturbation theory \cite{Palenik2016} so that equal occupation of all Fermi-level orbitals is no longer a requirement.  As before, the perturbed eigenvalues do not split.  Now, however, the unitary transformations that mix degenerate orbitals subdivide into two portions: one in which the occupation numbers are equal and one in which the occupation numbers are different.  The first-order perturbing potential can be diagonalized within each set of equally occupied orbitals by a zeroth-order unitary transformation.  The remaining off-diagonal matrix elements are zeroed by a combination of a first-order unitary transformation and change in first-order occupation numbers, which simultaneously restore degeneracy.

The perturbed quantities were generated by differentiating the unperturbed state with respect to the strength of an external perturbation.  The range over which these derivatives can be used to form a Taylor series connecting the perturbed and unperturbed states is limited, in part, by Fermi statistics.  When the perturbation grows large enough that the occupation numbers can no longer change, a derivative discontinuity will occur.

\begin{acknowledgments}
This work is supported by the Office of Naval Research, directly and through the Naval Research Laboratory. M.C.P. gratefully acknowledges an NRC/NRL Postdoctoral Research Associateship.
\end{acknowledgments}

\bibliography{citations}{}
\end{document}